# Effects of the removal of Ta capping layer on the magnetization dynamics of Permalloy thin films

Nikita Porwal[1], Simone Peli[2], P. J. Chen[3], Robert Shull[3] and Prasanta Kumar Datta[1*]

[1]*Department of Physics, Indian Institute of Technology Kharagpur, W.B. 721302, India*
[2]*Department of Physics, Indian Institute of Technology, Kharagpur, India*
*Current affiliation: Elettra-Sincrotrone Trieste S.C.P.A., 34149 Basovizza, Italy*
[3]*Materials Science and Engineering Division, National Institute of Standards and Technology, Gaithersburg, MD 20899, USA*
*Email: pkdatta@phy.iitkgp.ac.in

**Abstract**
We have investigated the spin wave dynamics of Permalloy (Py) thin films with and without a Ta capping layer for varying Py thickness (15 nm, 20 nm and 30 nm) using all optical time-resolved magneto-optical Kerr effect measurements. XPS measurements confirm the oxidation of the originally-prepared samples and also that the removal of the Ta capping layer is achievable by a few sputtering cycles. The magnetic field strength dependencies of the spin wave modes with the variation of the Py film thickness for the samples are studied. We observe that the presence of the Ta capping layer reduces the precessional frequencies of the samples while the samples without a Ta capping layer enhance the role of Py thickness. We also observe that the decay time of spin waves is highly dependent on the top layer of the samples. The decay time increases with increasing Py thicknesses for Ta/Py/Ta samples implying that the enhancement of decay time is caused by the Ta/Py/Ta interfaces. Whereas, for Ta/Py samples the decay time decreases with increasing Py thickness. The results of this work extend the knowledge on the magnetization dynamics of Py thin films giving information on how to resume and even enhance the spin mobility after a deleterious oxidation process. This can open new scenarios on the building process and on the maintenance of fast magnetic switching devices.

**Introduction**
Even though new data storage techniques (Solid State Disc) are being implemented, the magnetic based storage disks (Hard Disk Drives) are still the most prevalent way to keep data in memory and it seems that, in the nearest future, exploitation of the magnetic properties of ferromagnetic materials will be the key to develop faster and more capacious storage devices (MRAM) [1]. Elucidating the switching mechanisms and the switching speeds of the magnetic domains that occurs on picosecond (ps) or even femtosecond (fs) time scales is a crucial point for the realization of new switching devices. Such studies pose new questions, for example, about the temporal limits for the magnetization reversal process induced by external field pulses of short duration, and on the role of the damping parameter, the magnetization and magnetic anisotropies, and the system shape and size. For these reasons, specific experiments that can observe the movement of the spin domains are required. Ferromagnetic resonance (FMR) exploits the absorption of a microwave field by a material under the influence of an oscillating magnetic field [2, 3] and it is one of the most common techniques for deducing the precession frequency of the magnetic spins in a material. Time-resolve Magneto Optical Kerr Effect (TR- MOKE) [4] is, instead, a more direct technique for probing the relaxation dynamics of magnetic domains excited by an optical pulse. Since this precessional motion is captured directly in the time domain, it is possible to follow the evolution of the precession amplitude and so extract also its damping parameter. The TR-MOKE technique has other benefits as follows: (1) using short laser pulses, the temporal resolution is limited only by the width of such pulses (few fs); (2) exploiting the different magneto-optical configurations (polar, longitudinal and transverse) a spatial evolution of the spin excitation can be also achieved; (3) by the use of microscopic objectives one can have information about the local magnetization with a spatial resolution determined by the wavelength of the probe pulse through the Abbe diffraction limit [5]. In recent years, the scientific community has shown how multilayer films can be a good way toward achieving the controlled manipulation of the magnetization and the magnetic damping of several ferromagnetic materials. It has been shown how to make multilayer materials with tunable magnetization and damping [6, 7], and how the spin Hall effect may convert charge current to pure spin current in materials with large spin-orbit coupling [8]. Along with that, the presence of a thin non-magnetic capping layer can affect the damping parameter [9] and the coercivity [8] in Permalloy ($Ni_{80}Fe_{20}$, hereafter referred to as Py) thin films and can generate effects like spin pumping at the interfaces [4, 10]. In this context, transition metals like Tantalum (Ta) are widely used as seed and cap layers in the fabrication of spintronic devices like giant magneto-resistance effect spin valves, magnetic sensors and spin torque oscillators. It has been reported that Ta enables tuning of the post switching precessional motion of the magnetization of ferromagnetic materials [11]. As a consequence, the structural and magnetic properties of various kinds of Py/Ta structures have been studied [12-18]. It was found experimentally [14, 17] and later verified theoretically [14, 15, 19] that Ta seed layer intermixing produces a loss of moment equivalent to magnetically dead layers of 0.6-1.2 nm thickness and that a Ta cap layer results in a 1.06-0.2 nm thick dead layer. As the seed and cap layer interfaces are not chemically equivalent, it is not surprising their dead layer thicknesses are different. It has also been seen that 12% of Ta intermixing is sufficient to suppress the magnetic moments of the Ni atoms in Py thin films [14]. These dead layers constrain the propagation of spin waves (SW) in Py/Ta films and are thereby responsible for resulting in higher damping than that of Py thin films without a capping layer. Despite the large number of reports on the effects of Py/Ta capping studied by using FMR [20-24], few reports exist on the magnetization dynamics of Py/Ta thin films (and their corresponding damping characteristics) with varying Py thickness using Time-Resolved MOKE [25]. In this paper we measured the effect of a Ta capping layer on the magnetic precessional modes of a thin film of Py having three different thicknesses 15 nm, 20 nm and 30 nm by means of TR-MOKE. We have used femtosecond amplified laser pulses for excitation and detection of ultrafast magnetization dynamics in the samples. In particular, we investigated the precessional dynamics of the magnetic domains in the Py film excited by an ultrashort laser pulse before and after the removal of the Ta capping layer by means of a few sputtering cycles in an XPS chamber. The results show that the presence of the capping layer reduces the precessional frequencies of the spins while the absence of the Ta capping layer enhances the role of the Py thickness.

**Target and Methods**
Three sets of layered thin films Si/SiO$_2$(20)/Ta(4)/Py(t = 15, 20, and 30)/Ta (10) (thicknesses are in nm) were prepared by magnetron sputtering. Their schematic structure is shown in fig. 1(a). At first, a 20 nm thick SiO$_2$ film is grown on top of a silicon (Si) [001] substrate and then Ta(4)/Py(t = 15, 20, and 30)/Ta(5) were deposited using DC magnetron sputtering in an ultrahigh vacuum chamber at a base pressure of 6.6×10$^{-7}$ Pa (5×10$^{-9}$ torr), Ar pressure of 0.239 Pa (1.8 mtorr) at a power of 40 W. The capping layer of 5 nm Ta is usually used to avoid degradation from natural oxidation and exposure to a high-power laser during optical pump-probe experiments in air.

We investigate the SW dynamics of the thin film samples with varying in-plane bias magnetic field using TR-MOKE. The experiment was performed with a conventional two-color method and a standard all-optical pump-probe set-up. The interaction of the electromagnetic field of the light with the surface of the magnetized materials results in the rotation of the polarization of the incoming light. In the particular case of a reflecting material this effect is called the magneto optical Kerr effect (MOKE). The application of an external magnetic field $H$ on a ferromagnetic material generates a net magnetization M due to the alignment of the magnetic domains along the direction of the external field. If $H$ is time independent, $M$ is fixed, and the rotation of the polarization of an interacting light will be constant. MOKE is measured by comparing the directions of the polarization of the light before and after the interaction with the magnetic material. Figure 1(b) shows a schematic illustration of the principles of the TR-MOKE experiment. TR-MOKE essentially follows the dynamical change in the net magnetization M of the material initiated by an external excitation. Usually, the excitation is achieved by the means of a strong laser pulse with very short time duration. In this study, the ultrafast excitation pump ($\lambda$pump = 615 nm, repetition rate = 1 kHz, fluence = 10 mJ/cm$^{-2}$, pulse width = 100 fs) is provided by the OPA (Tomas prime) seeded by an amplified laser pulse while the fundamental beam ($\lambda_{probe}$= 808 nm, fluence = 2 mJ/cm$^{-2}$, pulse width = 50 fs) is used as a probe to detect the time varying polar Kerr rotation from the sample. A delay stage situated in the pump path is used to create the necessary time delay between the pump beam and the probe beam with temporal resolution of about 100 fs limited by the cross-correlation of the pump and the probe beams. Finally, both the beams are combined together and focused at the sample by a lens (focal length 300 mm)

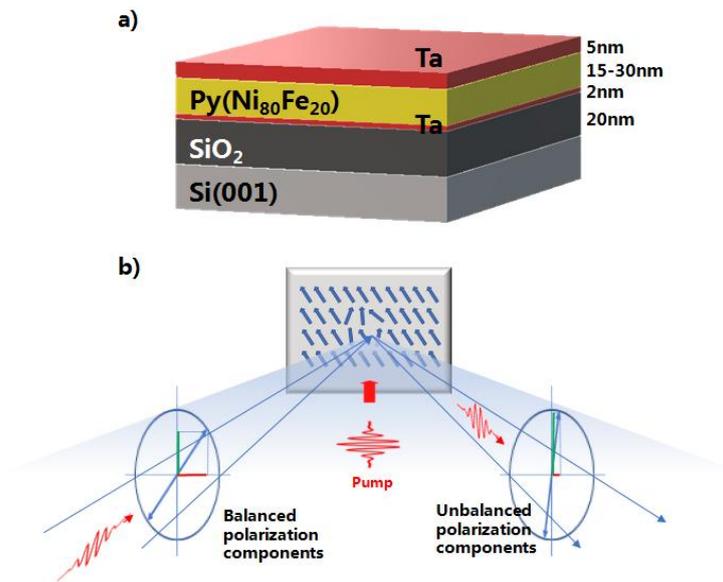

Figure1: (a) Sample consists of Si/SiO$_2$(20nm)/Ta(2nm)/Py(15-30nm)/Ta(5nm). Ta is used as seed and capping layer of the sample. (b) Shows schematic diagram of the polar Tr-MOKE. The balanced polarization state of incident light and after the reflection from the perturbed magnetized sample (unbalanced polarized light) are shown.

in a collinear geometry. The measured probe beam size is about 30 μm diameter while the size of the pump beam is 60 μm diameter. Under this condition the probe collects the dynamics from the uniformly excited part of the sample. The sample is placed in a controllable x-y-z stage to place the laser beams on different portions of the sample surface. An optical camera is used to provide an image of the two pump and probe beams on the sample so as to achieve a good spatial overlap. The MOKE geometry we adopted for this experiment is the polar configuration. In the polar configuration, the bias magnetic field $H$ is directed perpendicular to the sample surface and parallel to the plane of incidence of the laser light. Nevertheless, we applied the static magnetic field at a small angle (~15°) to the sample plane, the in-plane component of which is defined as the bias field $H$. The tilt of magnetization from the sample plane ensures a finite demagnetizing field along the direction of the pump beam, which is further modified by the pump pulse to induce precessional dynamics within the sample. A magnetized bar is placed on a translational stage and the magnitude of the in-plane component of this field is modified by moving the magnetic bar closer or farther away from the sample. The time varying polar Kerr rotation is measured at room temperature by using an optical bridge detector and lock-in amplifier in a phase sensitive manner. The pump beam is modulated at 400 Hz frequency which eventually is being used as the reference frequency of the lock-in amplifier. The total delay time window investigated is of about 1.3 ns and is sufficient to resolve the SW frequencies for the samples. All the experiments are performed under ambient condition and room temperature.

The alignment of the magnetic domains under the effect of an external constant magnetic field $H(t)$ can be disordered by means of a strong laser pulse. After very fast processes in the femtoseconds time-scale the dynamics of the total magnetization $M(t)$ is regulated by the

Landau-Lifshitz-Gilbert equation

$$\frac{dM}{dt} = -\gamma(M \times H_{eff}) + \alpha\left(M \times \frac{dM}{dt}\right)$$

that describes a damped precessional motion. Here, γ = gyromagnetic ratio and α is the damping constant. We follow this dynamical behavior by measuring the rotation of the polarization of the light when reflected by the magnetic material. After applying a static external magnetic field we separate the two components of the polarization of the reflected light using a Wollaston prism, and we rotate the prism in order to have the horizontal and the vertical components with exactly the same intensity. This intensity is read by two separate photodetectors. The Wollaston prism and the two photodetectors are incorporated in a single rotating optical bridge detector (OBD) that gives an output proportional to the difference between the reads from the photodetectors. In a static condition this output is zero. After switching on the optical pump, the imbalances between the horizontal and vertical components of the light polarization caused by the induced demagnetization are translated in a non-zero output from the OBD.

**Sample characterization**
We measured first the freshly prepared samples with the Ta capping layer on top of them. Some days after those measurements we found a complete disappearance of the MOKE signal on those same samples. We suspected that this degradation was provoked by oxidation. From the optical point of view oxidation can abruptly affect the skin depth of the light inside the material preventing both the pump and the probe to access the magnetic layer. From the magnetic point of view, oxidation can induce a modification of the hysteresis loop of the magnetic material modifying the saturation properties of the magnetization [26]. We used XPS to sputter the samples and check for the presence of oxygen on the surface of them. We have sputtered all three samples Si/SiO$_2$/Ta (2)/Py (t = 15, 20, and 30)/Ta (5) calling as A (Py thickness 15 nm), B (Py thickness 20 nm), and C (Py thickness 30 nm). Figure 2(a) shows Fe, Ni, Ta, O$_2$ and C contents at different depths of the sample C. Here, region I represents initial concentrations (i.e., data measured at 0 sputter cycle), region II shows data after 1/3$^{rd}$ of the sputtering cycle and region III represents the region where the Ta content is almost completely etched away and Ni and Fe contents reach maximum counts. With the increase of etching depth (region I, II and III of fig. 2 (a)), the changes in the XPS spectra for Fe, Ni, Ta and O$_2$ are shown in Fig 2 (b)-(e). For the Py/Ta bilayer, since the Ta layer on Py is thick (~5 nm), the peaks corresponding to Ta can be observed and underlying Py layer peaks (i.e. Ni or Fe peaks) cannot be observed in the surface analysis technique of XPS (region I). The high intensity Ta peaks at 22.2 eV and 24.0 eV (see fig. 2 (d)) observed, are characteristics of metallic Ta 4f$_{7/2}$ and Ta 4f$_{5/2}$ peaks respectively from the XPS handbook [27]. A small peak at 26.4 eV corresponds to Ta 4f$_{7/2}$ in Ta$_2$O$_5$. Ta$_2$O$_5$ was formed in the air during the experiment. Along with the Ta peaks, we also observed an O$_2$ peak in region I, at 530.7 eV, proving that the sample may be oxidized in the air at the time of the first pump-probe measurements. Thus, to get rid of the Ta capping layer, we etched away the top layer of the sample by using XPS sputtering. We observe that with increasing the etching depth of sample, the Ni and Fe contents are increasing and thus we observe small peaks are coming into the spectra. However, the Ta and $O_2$ contents are decreasing. The Ni 2p and Fe 2p peaks at region II are shown in fig. 2(b) and 2(c). Peak 1 at 853.0 eV and peak 2 at 870.3 eV (region II) are characteristic of metallic Ni $2p_{3/2}$ and $2p_{1/2}$ peaks respectively and peak 1 of Fe at 706.9 eV and peak 2 at 712.0 eV are not the metal Fe $2p_{3/2}$ and $2p_{1/2}$ peaks respectively. The Fe and Ni peaks may correspond to the intermetallic state of the sample due to the reaction between Ta, Ni and Fe at the interface of the Py/Ta sample. On further increase of etching rate, we observe that the Ni and Fe content is approaching a maximum and Ta and $O_2$ are approaching minimum values. At region III, the Ni content shows two peaks at 852.6 eV and 870.0 eV,

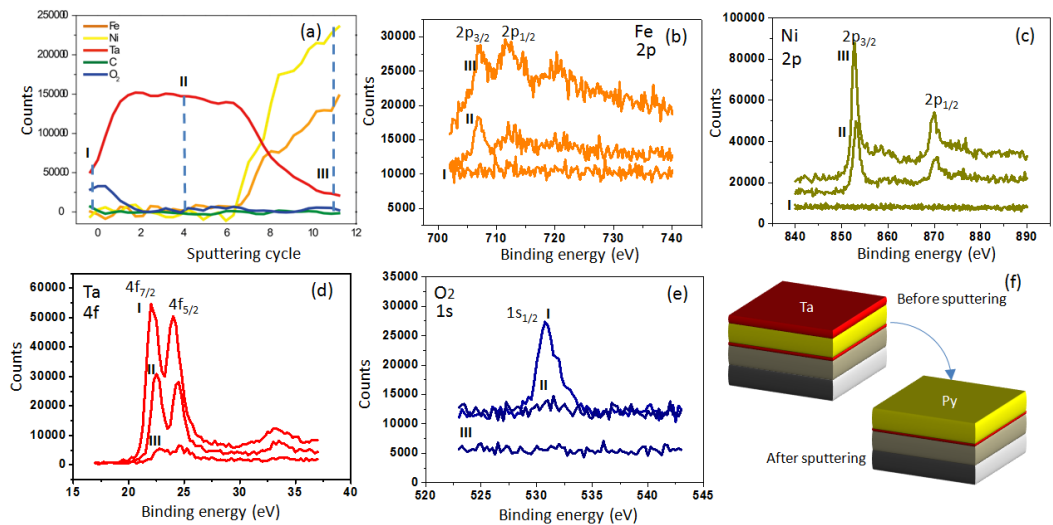

Figure2: (a) Variation of relative atom contents of iron, nickel, tantalum, carbon and oxygen with sputtering cycle for sample A. The dotted blue lines at I, II and III show the different sputtered regions of the sample. The XPS spectra of (b) Fe 2p, (c) Ni 2p, (d) Ta 4f and (e) O2 1s states show different sputtered regions of the sample A. (f) represent the schematic diagram of the sample before and after sputtering.

which correspond to the $2p_{3/2}$ and $2p_{1/2}$ pure metal Ni peaks respectively. The Fe content shows two prominent peaks at 706.8 eV and 711.8 eV which correspond to pure Fe metal peaks of $2p_{3/2}$ and $2p_{1/2}$ state respectively. We observe that the Ta peaks are reduced significantly and shifted to higher binding energy side with binding energy of 22.7 eV and 24.5 eV, which correspond to Ta $4f_{7/2}$ and $4f_{5/2}$ states which is not pure metallic in nature. The $O_2$ peak is no longer there in the spectra and thus reflects the non-oxidized state of the sample. We have sputtered a portion of each samples to have a complete characterization of the layers and a hint on the degree of oxidation. Then we moved on to a different portion and sputtered the surface for a fixed number of cycles just to completely remove the Ta capping layer while avoiding the removal of any of the Py layer. After the process of sputtering, the MOKE signal was recovered on all the samples. Thus, this procedure enabled us to do a comparison of the magnetization dynamics of Py films with and without the presence of the Ta capping layer.

## Results

Time-resolved Kerr rotation data for five different bias fields of sample C are presented in fig. 3(a) non-sputtered and (c) after sputtering, these data are obtained after subtracting the bi-exponential background. The background comes from the energy dissipation from the electron spins to the lattice and then from the lattice to the surroundings during the remagnetization process followed by the ultrafast demagnetization. Figures 3(b) and (d) show SW spectra obtained after performing the fast Fourier transform (FFT) of the non-sputtered sample and sputtered sample respectively. We observe that the uniform precessional mode frequency shifts to lower frequencies with decreasing bias magnetic field values, and this confirms the magnetic origin of the modes. The bias field dependence of the uniform precessional SW mode for the 15 nm, 20 nm and 30 nm Py thin film samples with Ta capping and without Ta capping is summarized in fig. 4.

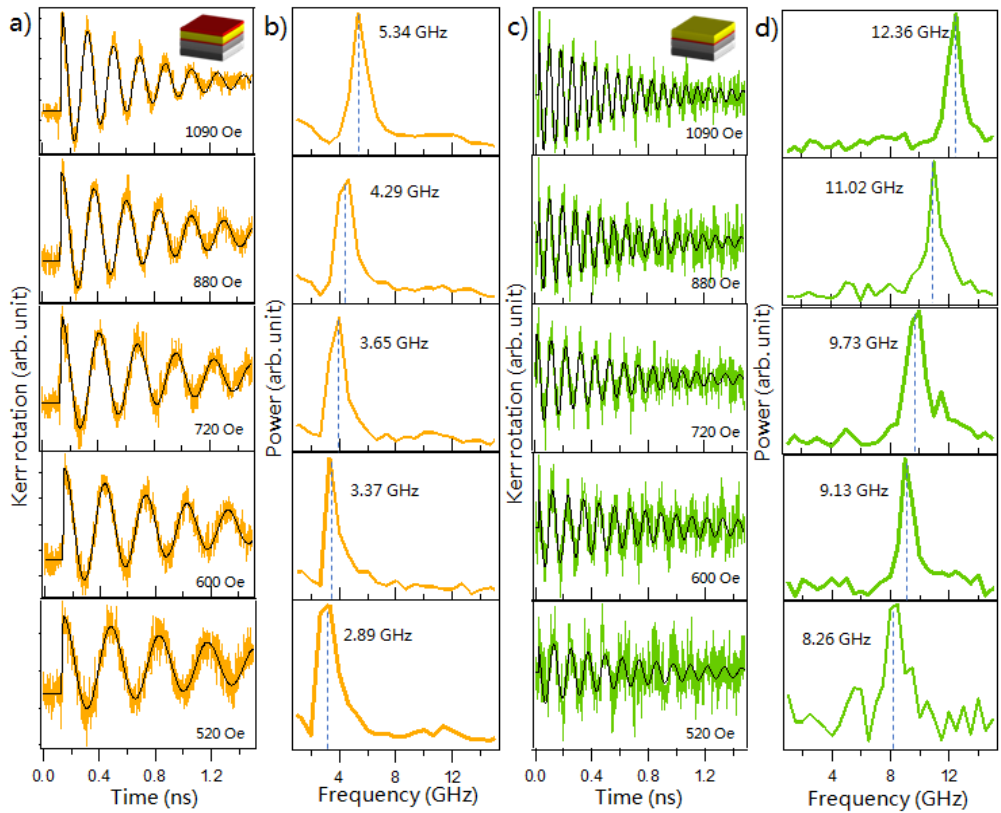

Figure 3: The background subtracted Kerr rotation data of (a) non-sputtered sample and (b) sputtered sample and corresponding fast Fourier transform spectra of (b) non-sputtered and (d) sputtered sample with varying bias magnetic fields are shown. The black lines in a) and c) are the fitted data with a single-mode damped oscillation.

There are two clear observations as follows: i) as the Py thickness increases, the shift in the frequency of the SW mode in the samples having a Ta cap (fig 4(a)) is very small while a significant shift in the SW frequency is observed for samples without the Ta capping layer (fig 4(b)). This may be due to the spin orbit interaction at the Py/Ta interface which suppresses the flow of spin in the sample and thus reduces the spin precession. (ii) The uniform precession SW modes for both the sputtered and non-sputtered samples show a monotonic increase in frequency with increasing bias magnetic field value. For uniform precession of magnetization, these uniform modes are fitted (black lines in fig. 4) with the Kittel formula as given by

$$f = \frac{\gamma}{2\pi}\sqrt{(H_b + H_k)(H_b + H_k + 4\pi M_s)}$$

where $\gamma$ is the gyromagnetic ratio, $H_b$ is the bias field, $H_k$ is the magneto-crystalline anisotropy and $M_s$ is the saturation magnetization of the sample. By using basic parameters of Py as $\gamma = 2.2$ and $H_k = 0$, we obtained $M_s$ values from the Kittel fit for all the samples. Surprisingly, the $M_s$ value does not change with the Py thickness for the non-sputtered samples (i.e. those having a Ta capping layer) and show almost the same values of 500 emu/cc. However, there is a monotonous decrease in the $M_s$ value with decreasing Py thickness for the sputtered samples (i.e. those without a Ta capping layer) and the values obtained are 888 emu/cc, 800 emu/cc and 500 emu/cc for the 30 nm, 20 nm and 15 nm Py thicknesses respectively.

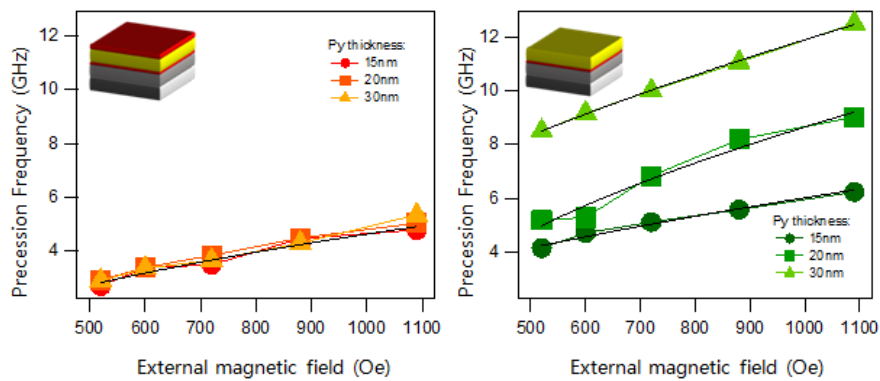

Figure 4: Precessional frequencies of different Py thicknesses of 15 nm, 20 nm and 30 nm for (a) non-sputtered sample and (b) sputtered samples are plotted as a function of bias magnetic fields. The insets show the schematic diagram of the non-sputtered and sputtered samples in (a) and (b) respectively. The black lines are the fitted curves with the Kittel function.

To extract the damping coefficient, the time-resolved Kerr rotation data at different bias magnetic fields are fitted with a damped sine curve given by

$$M(t) = M(0)e^{\frac{-t}{\tau}}sin(2\pi f t - \varphi)$$

The delay time $\tau$ is related to the damping coefficient $\alpha$ by the relation $\tau = 1/(2\pi f \alpha)$, where $f$ is the frequency of the uniform mode and $\varphi$ is the initial phase of the oscillation. The fitted data are shown by solid black lines in fig 2(a) and (c) for non-sputtered samples and sputtered samples respectively. The extracted values of $\tau$ for the Py films with Ta capping and without Ta capping layer are plotted with the external bias magnetic field (fig 5(a) and (b)) and with film thickness in (fig 5(c) and (d)) respectively. We observe that in the non-sputtered samples, the decay time decreases with increasing the bias magnetic field while for sputtered samples the decay time first decreases with increasing bias magnetic field and then at higher magnetic field it shows a sudden increase in decay time. However, decay time at fixed bias field increases with increasing the Py thicknesses for non-sputtered samples (see fig 5 (c)) while for sputtered samples (see Fig. 5d), the decay time decreases with increasing Py thickness. But a sudden increase in the decay time is observed for the 30 nm Py thick film.

**Discussion**

After the first measurements on fresh samples with fresh Ta capping, the oxygen deposited on the samples has given origin to a $Ta_2O_5$ layer which is responsible for increasing the extinction coefficient parameter. This prevents both the pump and probe wavelength light from penetrating the protective layer and exciting magnetization dynamics in the underneath magnetic layer. As a result, a complete suppression of the MOKE signal occurred. The presence of the oxide layer has been proven by XPS sputtering. After the removal of the oxide layer, the MOKE signal was again visible, but with some differences. The first big difference is the increase in the precessional frequencies in all three samples compared to the frequencies obtained for the samples with the capping layer. If we consider the case of the highest external magnetic field 1100 Oe, we have an increase in the precessional frequencies of 30%, 79% and 135% for the 15 nm, 20 nm and 30 nm thicknesses respectively. Another evidence coming from the experimental data is that in the case of the magnetic films covered by the Ta capping layer, the thickness of the Py layer only slightly affects the precessional frequencies as can be seen in fig. 4(a). After the removal of the capping layer the difference among the three Py thicknesses is much more evident. In fact, the precessional frequency increases with the thickness, confirming results already present in the literature [23]. This experimental data can be explained by the fact that the Py/Ta capping samples possess a spin transfer torque effect at the Py/Ta interface. It has been shown that currents through heavy metal layers, like Ta, can spin divide and create torques on adjacent ferromagnetic layers [24], thus suppressing the spin mobility and reducing the effect of the Py thickness. On the other hand, the SW frequency is higher for the sputtered thin film with respect to the non-sputtered one, which implies that the spin torque effect at the interface in the latter sample is not present due to not having a Ta capping layer.

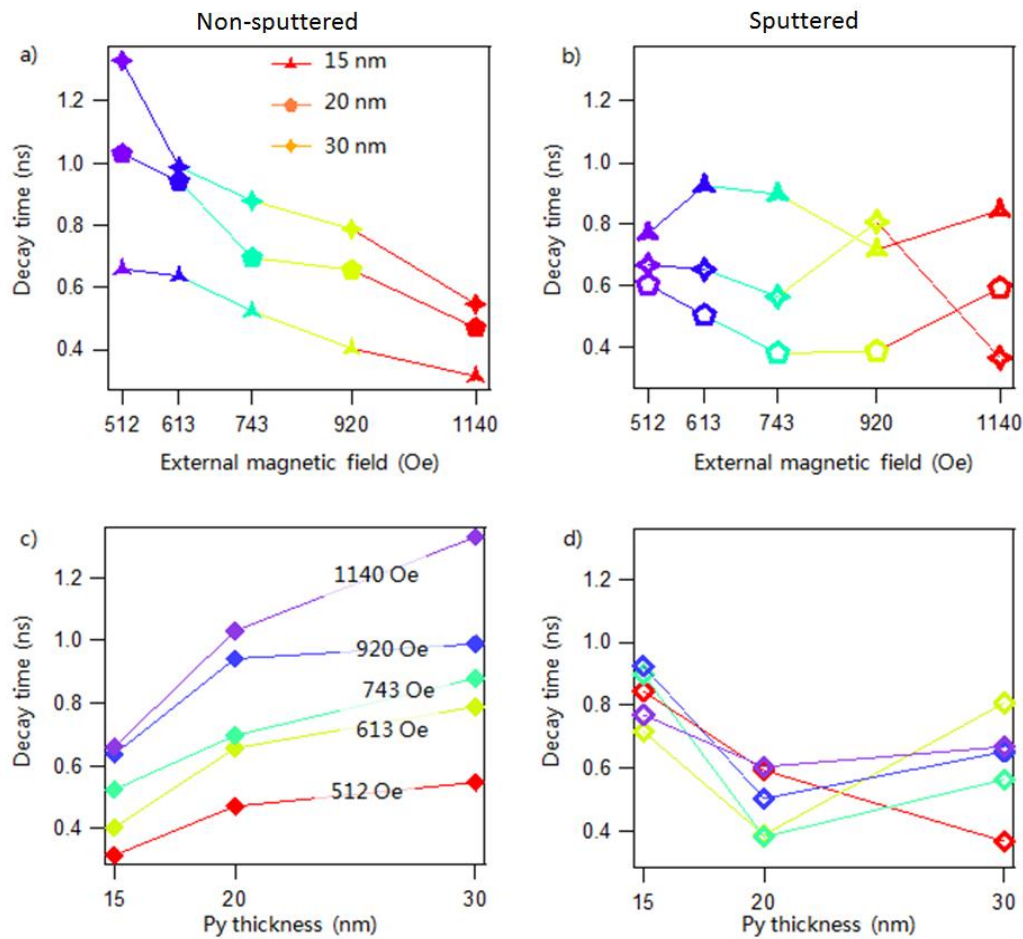

Figure 5. Decay time at various Py film thicknesses as a function of external bias magnetic field for (a) non-sputtered samples (solid symbols) and (b) for sputtered samples (open symbols) are plotted. (c) and (d) show, decal time at various external bias magnetic field as a function of Py thicknesses for non-sputtered (solid symbols) and sputtered samples (open symbols).

Interesting observations can be made from the damping data. On the fresh samples the dynamics decay time (inversely proportional to damping) shows a clear dependence with the thickness of the Py layer and with the external magnetic field. In particular, if we look at a single sample, the decay time decreases while the applied external magnetic field increases. This behaviour is confirmed for the all three samples (fig. 5(a)). If we fix the external magnetic field (fig. 5(c)) and change the Py thickness, we observe that the decay time increases together with the Py thickness. These systematic changes do not occur in the sputtered samples. In fact there is no clear monotonic dependency between decay times, Py thicknesses and external magnetic field. Nevertheless, it seems quite evident that the decay times tend to stay almost constant when the other parameters are changed and, overall, decay times are smaller than those of the non-sputtered samples. Theoretical explanation for the damping experimental data can be found in the fact that a spin current is generated by a precessing magnetization. In fact, magnetization dynamics of an ultrathin ferromagnetic layer sandwiched between two normal metal layers is also prone to interfacial effects such as the spin-pumping effect [24]. The spin current generated by a precessing magnetization can be absorbed by the normal metal in contact with the ferromagnetic layer, thus decreasing the damping (increasing the effective decay time) of the system.

**Conclusion**

The magnetization dynamics of both Ta/Py/Ta and Ta/Py films are investigated using an all-optical time resolved magneto-optical Kerr effect technique. Specifically, we have investigated the spin wave dynamics of the Py thin films with and without a Ta capping layer with varying Py thickness (15 nm, 20 nm and 30 nm). This study was triggered by the observation of the complete disappearance of the MOKE dynamics after a first cycle of measurements on the fresh samples. This disappearance was due to the oxidation of the Ta capping layer on the three samples. XPS data confirms oxidation of the originally-prepared samples and that the removal of the Ta capping layer can be achieved by a few sputtering cycles. We have observed significant variation in the precession frequencies with varying Py thickness for Ta/Py samples while the Ta/Py/Ta samples show almost the same frequencies with varying Py thicknesses. We have also observed that after Ta removal samples show completely reverse effects of the decay time than that of the samples having the Ta capping layer. The results of this work extend the knowledge on the magnetization dynamics of Py thin films giving information on how to resume and even enhance the spin mobility after a deleterious oxidation process. This can open new scenarios on the building process and on the maintenance of fast magnetic switching devices.